\documentclass[aps,prl,twocolumn,superscriptaddress,groupedaddress]{revtex4-1}
\usepackage{amsmath}
\usepackage{graphicx}
\usepackage{epstopdf}
\usepackage{bbold}
\usepackage{color}
\usepackage{float}
\usepackage{subfigure}
\usepackage[caption = false]{subfig}

\begin{document}
\title{Nanoscale Quantum Calorimetry with Electronic Temperature Fluctuations}
\author{F. Brange}
\affiliation{Department of Physics and NanoLund, Lund University, Box 188, SE-221 00 Lund, Sweden}
\author{P. Samuelsson}
\affiliation{Department of Physics and NanoLund, Lund University, Box 188, SE-221 00 Lund, Sweden}
\author{B. Karimi}
\affiliation{QTF Centre of Excellence, Department of Applied Physics, Aalto University, FI-000 76 Aalto, Finland}
\author{J. P. Pekola}
\affiliation{QTF Centre of Excellence, Department of Applied Physics, Aalto University, FI-000 76 Aalto, Finland}

\begin{abstract}
Motivated by the recent development of fast and ultra-sensitive thermometry in nanoscale systems, we investigate quantum calorimetric detection of individual heat pulses in the sub-meV energy range.  We propose a hybrid superconducting injector-calorimeter set-up, with the energy of injected pulses carried by tunneling electrons. Treating all heat transfer events microscopically, we analyse the statistics of the calorimeter temperature fluctuations and derive conditions for an accurate measurement of the heat pulse energies. Our results pave the way for novel, fundamental quantum thermodynamics experiments, including calorimetric detection of single microwave photons.
\end{abstract}

\maketitle

\emph{Introduction.---} In quantum calorimetry \cite{QuantCalPhysicsToday}, energy of individual particles is converted into a  
measurable temperature change. Mainly driven by the possibility of achieving unprecedented, high resolution and near-ideal efficiency x-ray detectors for space applications \cite{QuantCalPhysicsToday,MCCAMMON1993157,ApplPhysLett66.23.3203,doi:10.1063/1.1855411}, quantum calorimetry has over the past few decades also been developed for a wide range of other particles, including $\alpha$ and $\beta$ particles, heavy ions and weakly interacting elementary particles \cite{OVERLEY1985928,Nature314,CryoPartDet}. 
Today, fast and sensitive thermometry, together with small absorbers with weak thermal couplings to the surrounding, allows for time-resolved measurements \cite{doi:10.1063/1.1597983,PhysRevB.69.140301,PhysRevApplied.3.014007,Gov2016} and detection of energies all the way down to the far-infrared spectrum \cite{NatureWei,6005334}, i.e., energies of the order of meV. 

Recent demonstrations of fast and ultra-sensitive hot-electron thermometry \cite{PhysRevApplied.3.014007,Gov2016} at cryogenic conditions constitute a key step towards quantum calorimetry for even smaller energies, around 100 $\mu$eV or less. Time-resolved detection of such low-energy quanta, carried, e.g., by microwave photons or tunneling electrons, is of fundamental interest for nanoscale and quantum thermodynamics. This includes heat and work generation in open systems \cite{1367-2630-15-11-115006,1367-2630-16-11-115001,PhysRevE.90.022103,NaturePekola,PSSB:PSSB201600546}, thermodynamic fluctuation relations \cite{PhysRevLett.78.2690, PhysRevE.60.2721,PhysRevLett.102.210401,1751-8121-40-26-F08,PhysRevE.88.032146,PhysRevE.89.012127}, thermal quantum conductance \cite{NatureSchwab}, heat engines and information-to-work conversion \cite{NatureRio,NatureParrondo}, and coherence and entanglement \cite{PhysRevE.90.022103}. However, calorimetric sub-meV measurements still constitute an outstanding challenge; a proof-of-principle experiment requires an improvement of the detection sensitivity by at least an order of magnitude and a source of heat pulses with well defined energy and controllable injection rate. 

\begin{figure}[htb]
  \centering
 {\includegraphics[scale=0.46]{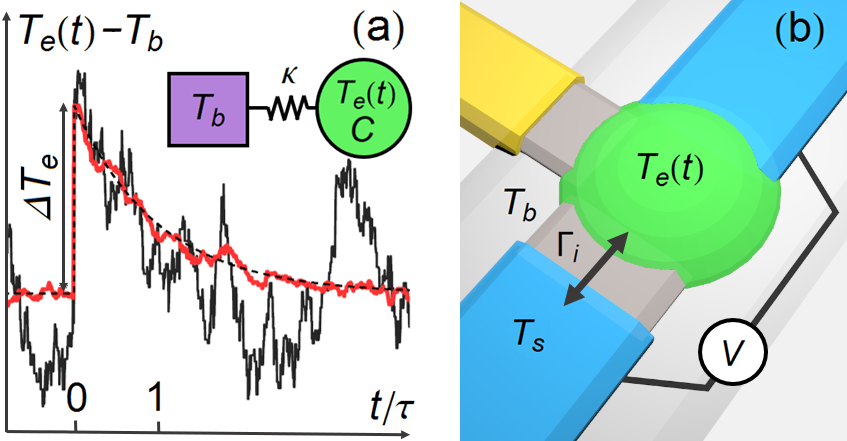}}
  \caption{(a) Two representative Monte Carlo simulated \cite{SI} time traces of the absorber electron temperature $T_\text e(t)$, with a jump $\Delta T_\text e$ caused by a single particle absorption event followed by a decay, rate $\tau$. The superimposed fluctuations are due to stochastic heat exchange with a phonon bath at low (red) and intermediate (black) temperatures $T_\text b$ (see text). Noise free case, Eq. (\ref{Absorber energy}), is shown with a dashed line. Inset: Effective circuit model of a calorimeter with heat capacity $C$ and heat conductance $\kappa$ to the bath. (b) Schematic of the nanoscale injector-calorimeter setup: A normal metallic island (green) contains a thermalized electron gas, with fluctuating temperature $T_\text e(t)$, constituting the absorber. The island is well coupled to an electrically grounded superconductor (upper, blue) acting as a heat mirror. It is further tunnel coupled to another superconductor (lower, blue), kept at a temperature $T_\text s$ and biased at a voltage $V$,  serving as a particle source with tunable injection rate $\Gamma_\text i(T_\text s,V)$. A thermometer, coupled to the island, is also shown (yellow). The island phonons, at temperature $T_\text b$, constitute a thermal bath weakly coupled to the island electron gas.}
\label{System}
\end{figure}

To meet this challenge, inspired by recent experiments \cite{PhysRevApplied.3.014007,Gov2016}, we propose and theoretically analyse a nanoscale hot-electron quantum calorimeter coupled to a superconducting injector, see Fig.~\ref{System}. The rate and energy of the injected heat pulses, carried by tunneling electrons, can be tuned by the applied injector bias and temperature. All calorimeter heat transfers, including the stochastic exchange of quanta with a weakly coupled thermal phonon bath, are treated on an equal, microscopic footing. By analysing the resulting calorimeter temperature fluctuations, focusing on the experimentally accessible lowest order cumulants, we derive conditions for a faithful operation of the calorimeter. Our results will stimulate novel, fundamental experiments, aiming for thermal measurements of, e.g., single microwave photons.

\emph{Hot-electron quantum calorimetry.---} A generic hot-electron quantum calorimeter is shown schematically in Fig.~\ref{System}~(a): An absorber with heat capacity $C$ is coupled, with thermal conductance $\kappa$, to a heat bath of phonons kept at temperature $T_\text{b}$. The absorber electron gas is rapidly thermalizing, with a temperature $T_\text e(t)$ well defined at all times. Operating in the linear regime and neglecting temperature background noise, absorbing a particle with energy $\varepsilon$ at $t=0$ gives rise to a jump $\Delta T_\text e=\varepsilon/C$ of the absorber temperature, followed by an exponential-in-time decay as 
\begin{equation}
T_\text e(t) =T_\text{b}+\Delta T_\text e e^{-t/\tau}, \qquad t\geq0
\label{Absorber energy}
\end{equation}
with $\tau = C/\kappa$ the relaxation time of the absorber. With a non-invasive and fast temperature measurement, $\Delta T_\text e$ and thus the energy $\varepsilon$ can be inferred. However, the background temperature exhibits fluctuations $\delta T_\text e(t)$, due to the fundamentally stochastic bath-absorber energy transfer, governed by the fluctuation-dissipation like relation 
\begin{equation}
\langle \delta T_\text e(t)\delta T_\text e(t')\rangle=\frac{k_\text B T_\text b^2}{C}e^{-|t-t'|/\tau},
\end{equation}
see Fig.~\ref{System} (a). Hence, the background noise can typically be neglected if the amplitude $\sqrt{\langle \delta T_\text e^2(t)\rangle}=T_\text{b}(k_\text B /C)^{1/2}$ is much smaller than the temperature signal $\Delta T_\text{e}$; larger noise prevents a faithful absorber temperature readout. 

The  condition  $\Delta T_\text{e} \gg \sqrt{\langle \delta T_\text e^2(t)\rangle}$ is met in state-of-the-art experiments \cite{PhysRevApplied.3.014007} with real-time detection of $\varepsilon \sim 100 $~meV, where the signal-to-noise ratio $\Delta T_\text{e}/\sqrt{\langle \delta T_\text e^2\rangle}=\varepsilon/[T_\text{b} \sqrt{k_\text B C}] \sim  100$ (for $T_\text{b}\sim 100$~mK, $C \sim 10^5 k_\text B$). To accurately detect $\varepsilon \lesssim 100~\mu$eV requires significantly reduced $C$ and $T_\text{b}$. While a standard dilution refrigerator reaches a temperature $\sim 10$  mK, careful design of the experiment is needed to reach that low $T_\text e(t)$. However, an equilibrium absorber electron temperature $\sim 30$ mK,  setting the effective bath temperature $T_\text b$, is fully feasible. Moreover, $C$ of a small metallic absorber at $T_\text b \sim 30$ mK can be as low as $\sim 10^3 k_{\rm B}$ \cite{PhysRevApplied.3.014007}, although some studies \cite{PhysRevB.97.115422} indicate that thin films exhibit higher values. The values $C\sim 10^3 k_{\rm B}, T_\text b=30$ mK yield a signal-to-noise ratio of order unity for an energy $\varepsilon \sim 100~\mu$eV, explicit absorber temperature time traces \cite{SI} with $\varepsilon = 200~\mu$eV and low, $T_\text b=5$~mK, and intermediate, $T_\text b=30$~mK, (signal-to-noise ratios $15$ and $2.4$ respectively) are given in Fig.~\ref{System}~(a) for reference. 

While these estimates show that a detection of heat pulses  $\varepsilon \lesssim 100 ~\mu$eV is within reach, albeit challenging, a proof-of-principle experiment also requires an injector with a controllable $\varepsilon$ and tunable injection rate $\Gamma_\text i$, such that the heat pulses are well separated in time, $\tau \Gamma_\text i \ll 1$. Here we propose and analyse an integrated hybrid superconductor injector-calorimeter, see Fig. \ref{System}, fullfilling all requirements. The injected heat pulses are carried by tunneling quasiparticles. Both the injector-absorber (i) and bath-absorber (b) heat exchanges are described microscopically, with quanta of energy transferred at rates  $\Gamma_\sigma(T_\text e)$, $\sigma=\text i, \text b$. The statistics of the heat pulses is described by the cumulant generating functions (CGFs) $F_{\sigma}(\xi_{\sigma},T_\text e)$ for the long-time, total energy transfer, as \cite{NJP2015}
\begin{equation}
F_{\sigma}(\xi_{\sigma},T_\text e) = \Gamma_{\sigma}(T_\text e) \left[\int d\varepsilon e^{i\varepsilon \xi_{\sigma}}P_{\sigma}(\varepsilon,T_\text e)-1 \right],
\label{poisson}
\end{equation}
for an uncorrelated, Poissonian, transfer of particles. Here $\xi_\text i,\xi_\text b$ are counting fields and the particle energies are distributed according to $P_\sigma(\varepsilon,T_\text e)$, accounting for fluctuations of energy due to quantum and/or thermal effects, generic for nanosystems. We first investigate the CGFs at constant $T_\text e$ and then analyse the back-action of the temperature fluctuations on the energy transfer rates, deriving estimates on the system parameters required for a faithful operation of the calorimeter. 

\emph{Hybrid nanoscale calorimeter.---} The injector-calorimeter system, shown in Fig. \ref{System} (b), consisting of a superconducting injector, with a gap $\Delta$ and kept at temperature $T_\text s$, is tunnel coupled, with a (normal state) conductance $G_T$, to a nanoscale metallic island absorber of volume $\mathcal{V}$. The absorber electron gas has a temperature $T_\text e(t)$ and a heat capacity $C[T_\text e(t)]=(\pi^2k_\text B^2/3)\nu_\text F T_\text e(t)$, where $\nu_\text F$ is the density of states at the Fermi level. The electron gas is further coupled \cite{PhysRevB.49.5942}, with a thermal conductance $\kappa[T_\text e(t)]=5\Sigma \mathcal{V}T_\text e^4(t)$ with $\kappa\equiv \kappa(T_\text b)$ and $\Sigma$ the electron-phonon coupling constant, to the bath phonons kept at a fixed temperature $T_\text b$. A second superconductor, coupled to the absorber island via an Ohmic contact, works as a heat mirror and fixes the  electric potential of the island to the superconducting chemical potential.  A bias voltage $V$, with $e|V|<\Delta$, is applied between the injector and the second superconductor. The temperature $T_\text e(t)$ is measured by a fast, ultra-sensitive thermometer, assumed to be effectively non-invasive \cite{Bayan}.  We also assume that both the standard and the inverse proximity effect can be neglected.  

Injector-absorber heat pulses are transferred by the tunneling of individual electron and hole quasiparticles. The statistical properties of the charge transfer across a normal-superconducting tunnel barrier are well known \cite{Muz1994,NazBook}. By properly accounting for the energy carried by each tunneling particle \cite{KinPil2004}, the generating function $F_\text i(\xi_\text i,T_\text e)$ for the heat transfer statistics is readily obtained as 
\begin{equation}
F_\text{i}(\xi_\text i,T_\text e)=\int d\varepsilon \left[\Gamma_+^\text i\left(e^{i\xi_\text i \varepsilon}-1\right) +\Gamma_ -^\text i\left(e^{-i\xi_\text i \varepsilon}-1\right)\right]
\label{scond}
\end{equation}
with rates $\Gamma_\pm^\text i(\varepsilon)=(G_\text T/e^2) \nu_\text S(\varepsilon-eV) f_\pm(\varepsilon-eV,T_\text s) f_{\mp}(\varepsilon,T_\text e)$  where $\nu_\text S(\varepsilon) = |\varepsilon|/\sqrt{\varepsilon^2-\Delta^2}\theta(|\varepsilon|-\Delta)$, with $\theta(\varepsilon)$ the step function, is the normalized superconducting density of states and $f_+(\varepsilon,T)=(e^{\varepsilon/[k_\text B T]}+1)^{-1}$, $f_-(\varepsilon,T)=1-f_+(\varepsilon,T)$. 
From the first and second derivatives of $F_\text{i}(\xi_\text i,T_\text e)$ with respect to $\xi_\text i$ (taken at $\xi_\text i \rightarrow 0$), the known expressions for the average energy current and noise \cite{GolKuz2001} are obtained. Equation \eqref{scond} describes particles tunneling in ($+$) and out ($-$) of the absorber with respective spectral rates $\Gamma_{\pm}(\varepsilon)$. The energy of each particle is "counted" via the factors $e^{\pm i \xi \varepsilon}$. By comparing Eqs.~\eqref{poisson} and  \eqref{scond} [changing $\varepsilon \rightarrow -\varepsilon$ in the second term in $\eqref{scond}$] we see that the injector provides uncorrelated-in-time energy transfer events, at a rate $\Gamma_\text i(T_\text e)=\int d\varepsilon \left[\Gamma_+^\text i(\varepsilon)+\Gamma_-^\text i(\varepsilon)\right]$, with an energy probability distribution $P_\text i(\varepsilon,T_\text e)=[\Gamma_+^\text i(\varepsilon)+\Gamma_-^\text i(-\varepsilon)]/\Gamma_\text i$. 

\begin{figure}[t]
  \centering {\includegraphics[scale=0.6]{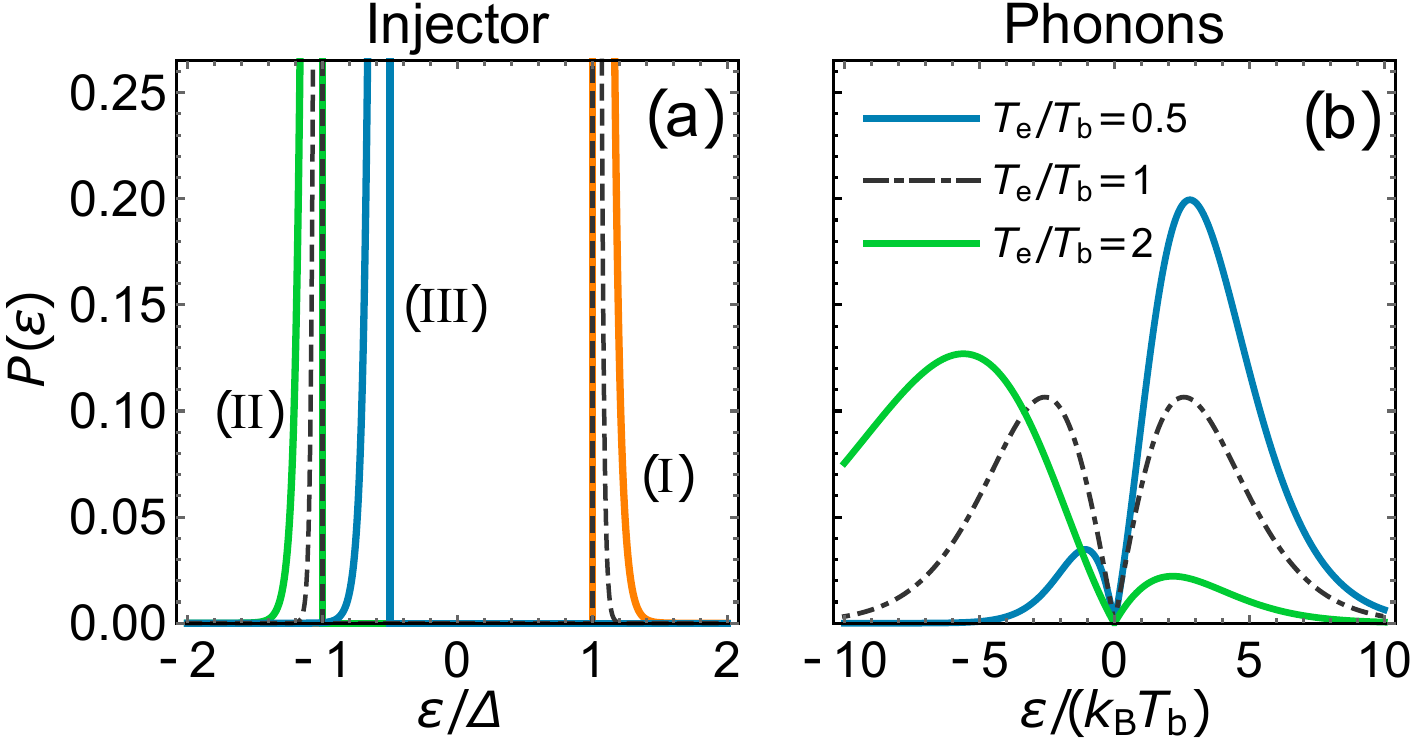}}
  \caption{(a) Probability distribution of energies transferred to the absorber $P(\varepsilon)$ from injector-absorber quasi-particle tunnelling, for four different sets of $\{k_\text B T_\text s/\Delta,k_\text B T_\text e/\Delta, eV/\Delta\}=\{0.02,0.02,0\}$ (dashed), $\{0.05,0.01,0\}$ (orange, solid), $\{0.01,0.05,0\}$ (green, solid) and $\{0.01,0.05,0.5\}$ (blue, solid). Corresponding injector regimes (I), (II) and (III) shown, see text. (b) Probability distribution for bath-absorber energy transfers due to phonon creation and annihilation, for different temperature ratios $T_\text e/T_\text b$.}
\label{probdist}
\end{figure}

Focusing on the regime $k_\text B T_\text s, k_\text BT_\text e \ll \Delta$, the CGF $F_\text{i}(\xi_\text i,T_\text e)$ describes four superimposed Poissonian processes with transfers at distinct energies $\pm \Delta\pm eV$ \cite{SI}.  In particular, in three different limits $V=0,T_\text s \gg T_\text e$ (I), $V=0,T_\text s \ll T_\text e$ (II) and $T_\text s(1 -e|V|/\Delta) \ll T_\text e \ll e|V|/k_\text B$ (III),  particles are injected at a corresponding energy $\varepsilon_\text I=\Delta$, $\varepsilon_{\text{II}}=-\Delta$ and $\varepsilon_{\text{III}}=eV-\Delta$, as clearly manifested in Fig.~\ref{probdist}~(a), giving CGFs
\begin{equation}
F_\text{i}^{(\alpha)}(\xi_\text i,T_\text e)=gc_{\alpha}\left(e^{i\varepsilon_{\alpha}\xi_\text i}-1\right), \quad \alpha=\text{I,II,III}
\label{CGFs}
\end{equation}
where $g=\sqrt{2\pi} G_\text T \Delta/e^2$ and $c_\text I=h(T_\text s)$, $c_\text{II}=h(T_\text e)$ and  $c_\text{III}=h(T_\text e)\exp([e|V|/k_\text BT_\text e)/2$, with  $h(T)=\sqrt{k_\text BT/\Delta}\exp(-\Delta/[k_\text B T])$. 

This analysis confirms that the superconductor constitutes a versatile injector, with particle energies and injection rates tunable via the externally controllable $T_\text s$ and $V$. Moreover, for small temperature deviations  $T_\text e-T_\text b \ll T_\text b$, relevant for the calorimeter operation, we have
\begin{equation}
\Gamma_\text i = g\left[h(T_\text s)+ h(T_\text b)\cosh\left(eV/k_\text B T_\text b\right)\right], 
\label{curr}
\end{equation}
Under the conditions  $C=10^3 k_{\rm  B},T_{\rm b}=30$~mK, the relaxation time $\tau$ is approximately 1-10~$\mu$s \cite{PhysRevApplied.3.014007,PhysRevB.97.115422}. For an aluminum superconductor with a gap $\Delta \approx 200~\mu$eV, the signal-to-noise ratio is 2.4. Experimentally $g\sim 10^{10}$-$10^{12}$ s$^{-1}$ if the injector resistance $G_\text T^{-1}$ varies in the range $3$-$300$ k$\Omega$ \cite{PhysRevApplied.3.014007,PhysRevB.97.115422}, making the individual injection event condition $\Gamma_\text i \tau \ll 1$ accessible by tuning $T_\text s,V$.  The superconducting injector is assumed to have ideal BCS (Bardeen-Cooper-Schrieffer) density of states (DOS). However, realistic tunnel junctions present non-zero leakage with zero-bias conductance $\gamma G_{\rm T}$ attributable to sub-gap states, absent in the BCS DOS. This leads to additional tunneling rate at sub-gap energies, $\Gamma_{\rm i}^0=\gamma g T_\text e/\Delta$, which however for standard $\gamma \sim~10^{-5}$ is negligible as compared to $\Gamma_{\rm i}$.

Microscopically, the bath-absorber energy transfer is due to creation and annihilation of individual bath phonons. Assuming a weak coupling between the phonons and the absorber electrons, the CGF $F_{\text b}(\xi,T_{\text e})$ of the energy transfer can be written in the form of Eq.~\eqref{scond}, with the spectral rates given by the text book result  \cite{Mahan} for phonons in a metal, $\Gamma_{\pm}^\text b(\varepsilon)= -\Sigma \mathcal{V}/[24k_\text B^5 \zeta(5)]\varepsilon^3 n(\pm \varepsilon,T_\text b)n(\mp \varepsilon,T_\text e)$, where $n(\varepsilon,T) = (e^{\varepsilon/[k_\text BT]}-1)^{-1}$ and $\zeta(x)$ the Riemann zeta function. Similar to the injector, from $\Gamma_{\pm}^\text b (\varepsilon)$ one gets  $\Gamma_\text b(T_\text e)=\int d\varepsilon \left[\Gamma_+^\text b(\varepsilon)+\Gamma_-^\text b(\varepsilon)\right]$ and $P_\text b(\varepsilon,T_\text e)=[\Gamma_+^\text b(\varepsilon)+\Gamma_-^\text b(-\varepsilon)]/\Gamma_\text b$, with the energy probability distribution plotted in Fig.~\ref{probdist}~(b) for a set of temperature ratios $T_\text e/T_\text b$. It is clear from the figure that, in contrast to the sharply peaked and gapped injector-absorber energy distribution, the bath-absorber distribution is broad and smooth, symmetric around $\varepsilon=0$ for $T_\text e=T_\text b$.  

The cumulants $S^{(n)}_\text b=\partial_{\xi_\text b}^nF_\text b(\xi_\text b,T_\text e)|_{\xi_\text b=0}$ are given by
\begin{equation}
S^{(n)}_\text b=\Sigma \mathcal{V}k_\text B^{n-1}\frac{\zeta(n_{\pm})(n+3)! }{24 \zeta(5)}\left(T_\text e^{n+4}\pm T_\text b^{n+4}\right),
\label{FCSb}
\end{equation}
where $n_\pm=n+(7 \pm 1)/2$ and $+/-$ is for $n=1,2...$ even/odd.  The result for odd $n$ is exact and for even $n$ an accurate approximation, deviating $<2\%$ from the exact result for any $n,T_\text e/T_\text b$ \cite{Pekola2018,SI}.  We note that $S^{(1)}_\text b=\Sigma \mathcal{V}(T_\text e^5-T_\text b^5)$, the well-known average bath-absorber energy current \cite{PhysRevB.49.5942}.

\emph{Temperature fluctuation statistics.---} While the average temperature in hybrid nanoscale systems has been widely investigated \cite{GiaPek2006}, there is to date no experimental investigation of the temperature noise. To obtain a complete picture of the fluctuations we investigate the full temperature statistics \cite{HeikNaz2009,LakHeikNaz2010,LakHeikNaz2012,Batt2013}, however the focus is on the noise, i.e., the second cumulant of the distribution. We note that both rates $\Gamma_\sigma(T_\text e)$ and probabilities $P_\sigma(\varepsilon, T_\text e)$ generally depend on the absorber temperature $T_\text e$. As a result of the stochastic energy transfers, $T_\text e(t)$ develops fluctuations in time, which in turn acts back on the transfer statistics. Fully accounting for this back-action effect, we analyse the distribution $P(\theta)$ of the low-frequency, time integrated absorber temperature fluctuations $\theta = \int [T_{\text e}(t)-\overline{T}_\text e] dt$, with $\overline{T}_\text e$ the average electron temperature. The $P(\theta)$ as well as the cumulants are obtained within a stochastic path integral approach \cite{JmatPhys2004}, following \cite{NJP2015}. 

\begin{figure}[htb]
  \centering {\includegraphics[scale=0.6]{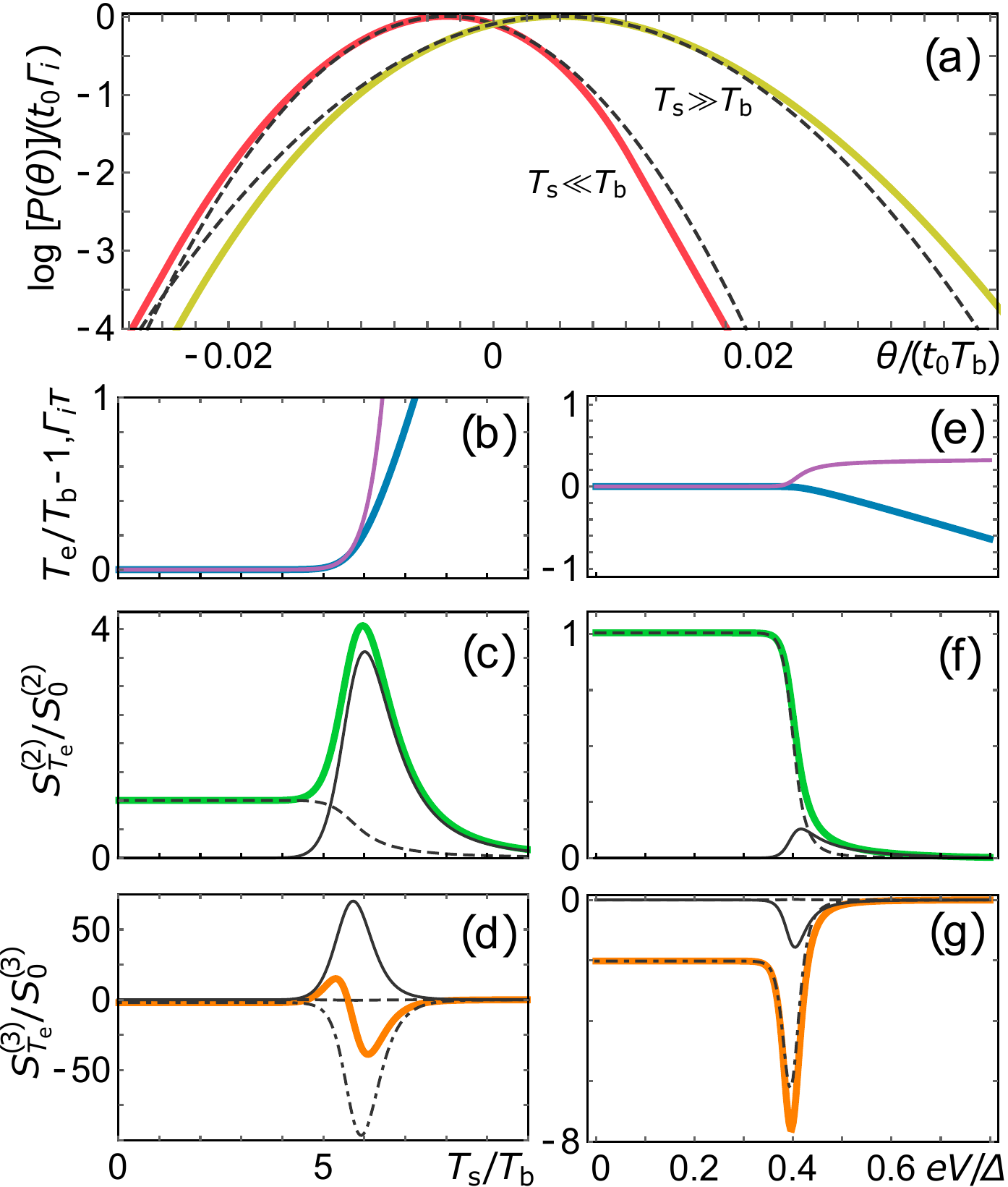}}
  \caption{(a) Temperature probability distribution $P(\theta)$ for injector parameters $T_\text s=10T_\text b$ (red solid line) and $T_\text s=0.1T_\text b$ (yellow, solid line), corresponding to injector cases (I) and (II) respectively. Dashed lines show the respective best Gaussian fits. In both plots $V=0$, $T_\text b=0.01 \Delta/k_\text B$, $C = 20\Delta/T_\text b$ and $\tau \Gamma_\text i=0.1$. (b)-(g) The first three cumulants as a function of $T_\text s/T_\text b$, at $V=0$ [(b) - (d)]  and $eV/\Delta$, at $T_\text s=T_\text b$ [(e) - (g)]. In all panels $T_\text b=0.01 \Delta/k_\text B$, $C = 20\Delta/T_\text b$. The total cumulants are shown with thick, solid lines in all panels. In (b) and (e), $\tau \Gamma_\text i$ is also shown (purple, thin solid line). In (c), (d), (f), (g) the injector-absorber (thin, solid line) and bath-absorber (thin, dashed line) contributions to the respective cumulants are shown. In (d) and (g) the back-action component (dash-dotted line) is shown.}
\label{tempstat}
\end{figure}

The distribution is plotted in Fig.~\ref{tempstat}~(a) for the two different regimes (I) and (II), with injection at energies $\pm \Delta$, at  $\tau \Gamma_\text i \ll 1$. As a consequence of the heat pulses being well separated in time, the deviations from the average $T_\text b t_\text 0$ are small ($t_0$ is the measurement time). However, the two distributions are clearly non-Gaussian, shifted and skewed in opposite temperature directions. The average electron temperature $\overline{T}_\text e$ as well as the cumulants  $S_{\text T_\text e}^{(n)}$ can be expressed in terms of $\langle \! \langle \mathcal{E}^n (T_\text e)\rangle \! \rangle=(-i)^n\partial_{\xi}^n F(\xi,T_\text e)|_{\xi=0}$, the cumulants of the absorber energy currents. Here $F(\xi,T_\text e) = F_\text i(\xi,T_\text e)+F_\text {b}(\xi,T_\text e)$. The average temperature $\overline T_\text e$ is found from the energy conservation condition 
\begin{equation}
\langle \mathcal{E} (\overline T_\text e)\rangle=0.
\end{equation}
The second cumulant, i.e., the temperature noise, and the third cumulant are given by \cite{SI} 
\begin{eqnarray}
\nonumber
S_{\text T_\text e}^{(2)}&=&\frac{1}{\kappa^2}\langle \! \langle \mathcal{E}^2(T_\text e) \rangle \! \rangle , \\
S_{\text T_\text e}^{(3)}&=&\frac{1}{\kappa^3}\left[\langle \! \langle \mathcal{E}^3(T_\text e) \rangle \! \rangle+3\langle \! \langle \mathcal{E}^2(T_\text e) \rangle \! \rangle \frac{d}{dT_\text e}\frac{ \langle \! \langle \mathcal{E}^2(T_\text e) \rangle \! \rangle}{\kappa (T_\text e)}\right],
\label{S2S3}
\end{eqnarray}
where the second term in $S_{\text T_\text e}^{(3)}$ is due to the back-action. In Eq. (\ref{S2S3}) $\kappa (T_\text e)=i\partial_{T_\text e}\partial_\xi F(\xi,T_\text e)|_{\xi=0}$, and all quantities are evaluated at $\overline T_\text e$. In the lower panels in Fig.~\ref{tempstat}, (b)-(g), $\overline T_\text e$, $S_{\text T_\text e}^{(2)}$ and $S_{\text T_\text e}^{(3)}$ are plotted for relevant parameters $C = 10^3 k_\text B$ and $T_\text b=0.01 \Delta/k_\text B$, i.e., $T_\text b \approx 20$~mK for an Al injector with $\Delta=200~\mu$eV. Two cases, thermal ($V=0$) and voltage ($T_\text s=T_\text b$) bias, are presented separately. 

\emph{Thermal bias.---} We focus on the experimentally relevant regime $\beta \gg \ln(r) \gg 1$, with $\beta=\Delta/(k_\text B T_\text b)$ and $r=g \Delta/[T_\text b \kappa]$.  Upon increasing $T_\text s$, the average temperature  $\overline T_\text e=T_\text b[1+5rh(T_\text s)]^{1/5}$ shows [Fig.~\ref{tempstat}~(b)] a cross-over at $T_\text s \sim T_\text s^*\equiv \Delta /[k_\text B \ln (r)]$ from constant, $T_\text b$ (dominated by bath coupling), to exponentially increasing $\sim e^{-\Delta/[5k_\text B T_\text s]}$ (dominated by injector coupling). The cross-over temperature sets the upper limit for operation of the calorimeter; since $T_\text s^* \gg T_\text b$ we are in the injector regime (I) , with a well defined particle energy $\Delta$. However, the rate $\Gamma_\text i \approx g h(T_\text s^*)\approx \kappa T_\text b/[\Delta \sqrt{\ln(r)}]$, giving $\tau \Gamma_i \approx C T_\text b /[\Delta \sqrt{\ln(r)}] \sim 1$ for relevant parameters.

The temperature fluctuations $S_{\text T_\text e}^{(2)}$, normalized to the equilibrium phonon noise $S_0^{(2)}=2k_\text BT_\text b^2/\kappa$, can be written as a sum of the bath and injector noise as
\begin{eqnarray}
S_{\text T_\text e}^{(2)}/S_{0}^{(2)}=\frac{1+q^6}{2q^8}+\frac{\beta (q^5-1)}{10 q^8},
\label{taunoise}
\end{eqnarray}
where $q \equiv \overline T_\text e/T_\text b$. As shown in Fig. \ref{tempstat} (c), upon increasing $T_\text s$ the bath noise decreases while the injector noise first increases. The total noise peaks at $T_\text s \approx T_\text s^*$ and then decays towards zero, due to the increase of the thermal conductivity $\kappa(\overline T_\text e)=\kappa q^4$. The peak value, to leading order in $1/\beta \ll 1$, is $S_{\text T_\text e}^{(2)}/S_{0}^{(2)}\approx 0.035 \beta$. Note that in the regime of optimal calorimeter operation, $\overline T_\text e-T_\text b\ll T_\text b$, we have  $S_{\text T_\text e}^{(2)}=S_{0}^{(2)}+ \Gamma_\text i(T_\text b) \langle \varepsilon^2 \rangle/\kappa^2$, i.e., by subtracting the equilibrium phonon noise, the second moment  $ \langle \varepsilon^2 \rangle$ of $P_\text i(\varepsilon,T_\text b)$, can be directly inferred. 

The third cumulant is plotted in Fig.~\ref{tempstat}~(d).  At low temperatures $T_\text s \ll T_\text s^*$ , $S_{\text T_\text e}^{(3)}$ is dominated by the back-action term, giving $S_{T_\text e}^{(3)}/S_{0}^{(3)}=-2$, with $S_{0}^{(3)}=6k_\text B^2T_\text b^3/\kappa^2$. Increasing $T_\text s$ the cumulant changes sign twice around $T_\text s^*$, a consequence of a competition between the positive injector term and the negative back-action term (the phonon contribution is negligibly small). 

\emph{Voltage bias.---}  The average temperature  $\overline T_\text e$ as a function of $V$ shows [Fig.~\ref{tempstat}~(e)] a cooling effect  \cite{GiaPek2006}, with a cross-over around $V\sim V^*=[\Delta-\ln(r) k_\text BT_\text b]/e$ from constant, $T_\text b$ to close-to-linear decrease $k_\text B \overline T_\text e \approx (\Delta-eV)/\ln(r)$. The cross-over voltage sets the upper limit for operation of the calorimeter since the condition $\Gamma_\text i \tau \ll 1$ breaks down for $V > V^*$. 

The normalized fluctuations can be written as a sum of the bath ($\propto 1+ q^6$) and injector ($\propto 1-q^5$) noise as, introducing $\tilde \beta=\beta(1-eV/\Delta)$, 
\begin{eqnarray}
\frac{S_{\text T_\text e}^{(2)}}{S_{0}^{(2)}}=\frac{q^4}{2}\frac{1+q^6+(\tilde \beta/5)(1-q^5)}{\left(q^6+(\tilde \beta/5)(1-q^5)\right)^2}.
\end{eqnarray}
As shown in Fig.~\ref{tempstat}~(f), at $V<V^*$, the noise is dominated by the (equilibrium) phonon part while for $V>V^*$ the noise decreases monotonically with increasing $V$, due to the increasing thermal conductivity $\kappa(\overline T_\text e)=\kappa(q^4+\tilde \beta(1-q^5)/[5q^2])$. The third cumulant  $S_{\text T_\text e}^{(3)}$ is dominated, for $V<V^*$, by the back-action term, giving $S_{T_\text e}^{(3)}/S_{0}^{(3)}=-2$. With increasing bias the cumulant first become increasingly negative, reaching a minimum around $V^*$ and thereafter decrease in absolute magnitude, towards zero, see Fig.~\ref{tempstat}~(g). Experimentally, a finite $V$ can lead to simultaneous changes of $T_\text e(t)$ and $T_\text s$, not discussed here.

\emph{Conclusions and outlook.---} We have proposed and theoretically analyzed nanoscale quantum calorimetry of individual tunnelling electrons in a hybrid superconducting set-up. We show that sub-meV calorimetry is feasible under optimized experimental conditions. The achievable signal-to-noise ratio is dictated by temperature fluctuations and backaction effects. Our results will spur advanced investigations of experimentally relevant phenomena such as the effect of a non-equilibrium electron distribution of the absorber and the invasive effect of the temperature measurement.  
  
\begin{acknowledgments}
\emph{Acknowledgements.---} We acknowledge discussions with V. Maisi and P. Hofer. F.B. and P.S. acknowledge support from the Swedish Research Council. This work was funded through Academy of Finland grant 312057 and from the European
Union's Horizon 2020 research and innovation programme under the European Research Council (ERC) programme
and Marie Sklodowska-Curie actions (grant agreements 742559 and 766025). 
\end{acknowledgments}

\bibliographystyle{apsrev4-1}
\bibliography{sources}

\end{document}


\begin{centering}
{\Large Supplemental Material} \vspace{3mm} \\ {\large Nanoscale Quantum Calorimetry with Electronic Temperature Fluctuations } \\ \vspace{3mm} F. Brange, P. Samuelsson, B. Karimi, and J. Pekola\\ $\vspace{5mm}$
\end{centering}

\subsection*{Monte Carlo simulations}
Here we present some examples of Monte Carlo generated time traces of the temperature fluctuations. The simulations are fully taking into account both the stochastic injector events, transferring energy according to the CGF in Eq.~(4) of the main text, and the stochastic phonon emission and absorption events. From the simulations we obtain numerical values of the average temperature, noise and skewness. Key expressions like Eqs.~(9), (10) and (11) of the main text have been found to be in perfect agreement with the Monte Carlo simulations. \\

\noindent  In Fig.~1, we show examples of time traces for $T_\text b = 5$~mK, $T_\text b=30$~mK and $T_\text b=100$~mK, respectively, to illustrate the effect of phonon noise at different temperatures. In all cases, $\varepsilon = 200$ $\mu$eV,  $C=1000k_\text B$ and time is chosen such that an injector event takes place at $t=0$. The three cases correspond to $\Delta T_\text e/\sqrt{\langle \delta T_\text e^2 \rangle} = 15, 2.4$ and $0.73$, respectively. As clearly seen, at low temperatures [see Fig.~1~(a)], the background noise is almost negligible compared to the temperature spike induced by the injector. For more experimentally realistic settings with intermediate temperatures [see Fig.~1~(b)], the temperature spike of the injector is still clearly visible, although the background noise is no longer negligible. At even higher temperatures [see Fig.~1~(c)], the temperature spike induced by the injector drowns in phonon noise and it gets difficult to identify the injector events.

\subsection*{Generating function for the injector-absorber energy transfer}
Here we derive the cumulant generating function for the superconducting injector given in Eq.~(5) of the main text. Our starting point is Eq.~(4) of the main text,
\begin{equation}
F_\text i(\xi_\text i,T_\text e) = \int d\varepsilon \left[ \Gamma_+^\text i \left( e^{i\xi_\text i \varepsilon}-1\right)+\Gamma_-^\text i \left( e^{-i\xi_\text i \varepsilon}-1\right) \right],
\label{FCS injector}
\end{equation}
with rates $\Gamma^\text i_\pm(\varepsilon) = \left( G_\text T/e^2\right)\nu_\text S(\varepsilon-eV)f_\pm(\varepsilon-eV,T_\text s)f_\mp(\varepsilon,T_\text e)$, where $\nu_\text S(\varepsilon) = |\varepsilon|/\sqrt{\varepsilon^2-\Delta^2}\theta(|\varepsilon|-\Delta)$ is the normalized superconducting density of state, $f_+(\varepsilon,T)=(e^{\varepsilon/[k_\text B T]}+1)^{-1}$ and $f_-(\varepsilon,T)=1-f_+(\varepsilon,T)$.\\

\noindent For $k_\text BT\ll \Delta-e|V|$, $T=T_\text s,T_\text e$, only the tails of the Fermi functions contribute to the integral. Eq.~\eqref{FCS injector} can then be written as
\begin{eqnarray}
\nonumber
F_\text i(\xi_\text i,T_\text e) &=& \frac{G_\text T}{e^2} \Bigg( \int_{\Delta+eV}^\infty d\varepsilon \frac{\varepsilon-eV}{\sqrt{(\varepsilon-eV)^2-\Delta^2}} \left[ e^{-(\varepsilon-eV)/[k_\text B T_\text s]} \left( e^{i\xi_\text i \varepsilon}-1\right)+e^{-\varepsilon/[k_\text BT_\text e]} \left( e^{-i\xi_\text i \varepsilon}-1\right) \right]\\
\nonumber
&&- \int^{-\Delta +eV}_{-\infty}d\varepsilon \frac{\varepsilon-eV}{\sqrt{(\varepsilon-eV)^2-\Delta^2}}\left[ e^{\varepsilon/[k_\text B T_\text e]} \left( e^{i\xi_\text i \varepsilon}-1\right)+e^{(\varepsilon-eV)/[k_\text BT_\text s]} \left( e^{-i\xi_\text i \varepsilon}-1\right) \right]  \Bigg)\\
\nonumber
&=& \frac{G_\text T}{e^2} \Bigg( \int_{\Delta+eV}^\infty d\varepsilon \frac{\varepsilon-eV}{\sqrt{(\varepsilon-eV)^2-\Delta^2}} \left[ e^{-(\varepsilon-eV)/[k_\text B T_\text s]} \left( e^{i\xi_\text i \varepsilon}-1\right)+e^{-\varepsilon/[k_\text BT_\text e]} \left( e^{-i\xi_\text i \varepsilon}-1\right) \right]\\
&&+ \int_{\Delta -eV}^{\infty}d\varepsilon \frac{\varepsilon+eV}{\sqrt{(\varepsilon+eV)^2-\Delta^2}}\left[ e^{-(\varepsilon+eV)/[k_\text BT_\text s]} \left( e^{i\xi_\text i \varepsilon}-1\right)+e^{-\varepsilon/[k_\text B T_\text e]} \left( e^{-i\xi_\text i \varepsilon}-1\right) \right]  \Bigg).
\end{eqnarray}

\begin{figure}[H]
  \centering {\includegraphics[scale=0.8]{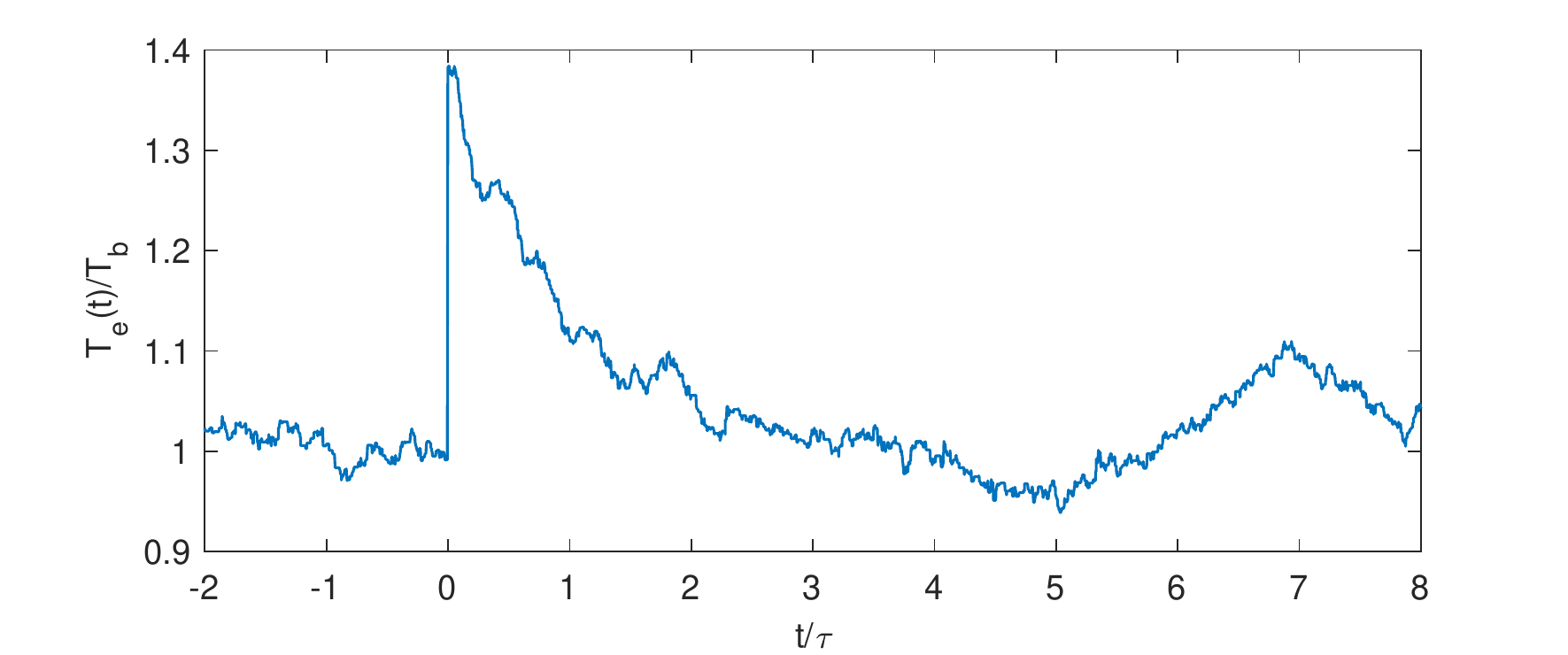}\quad\includegraphics[scale=0.8]{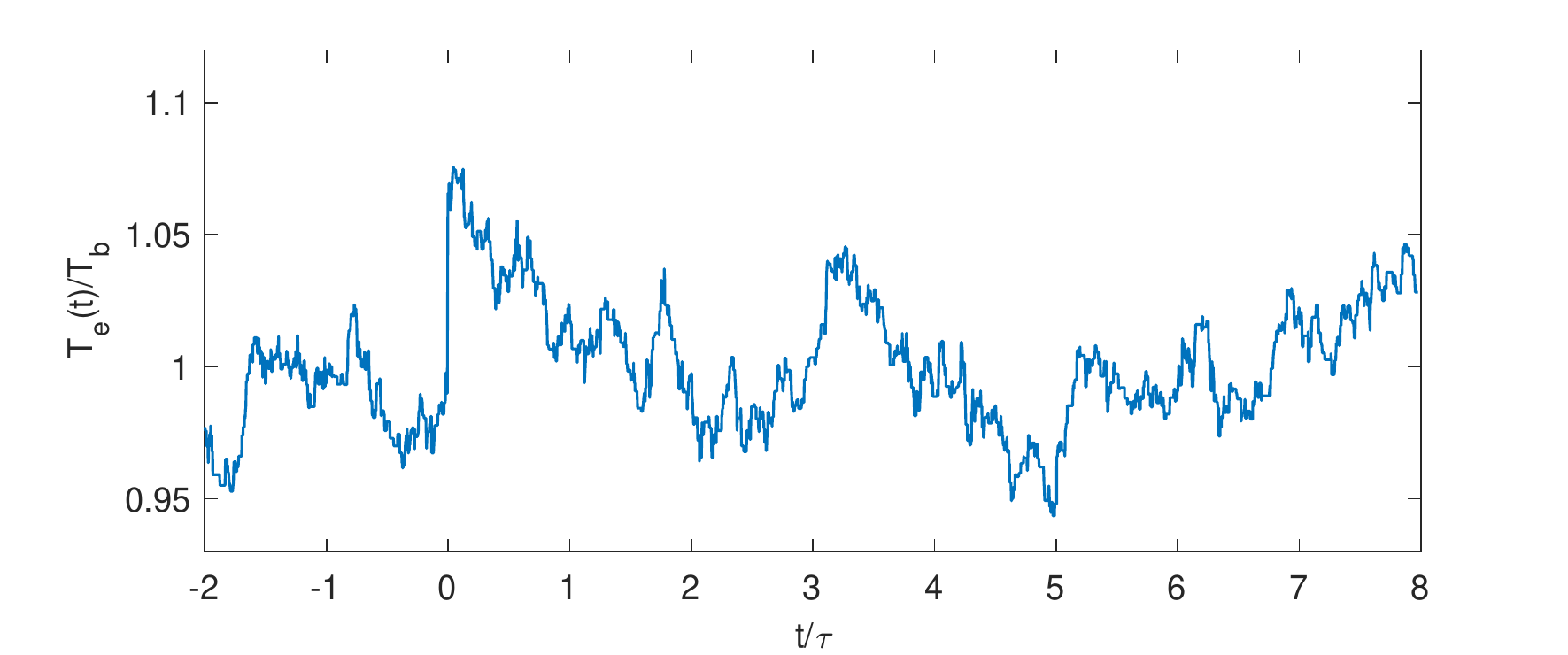}\quad \includegraphics[scale=0.8]{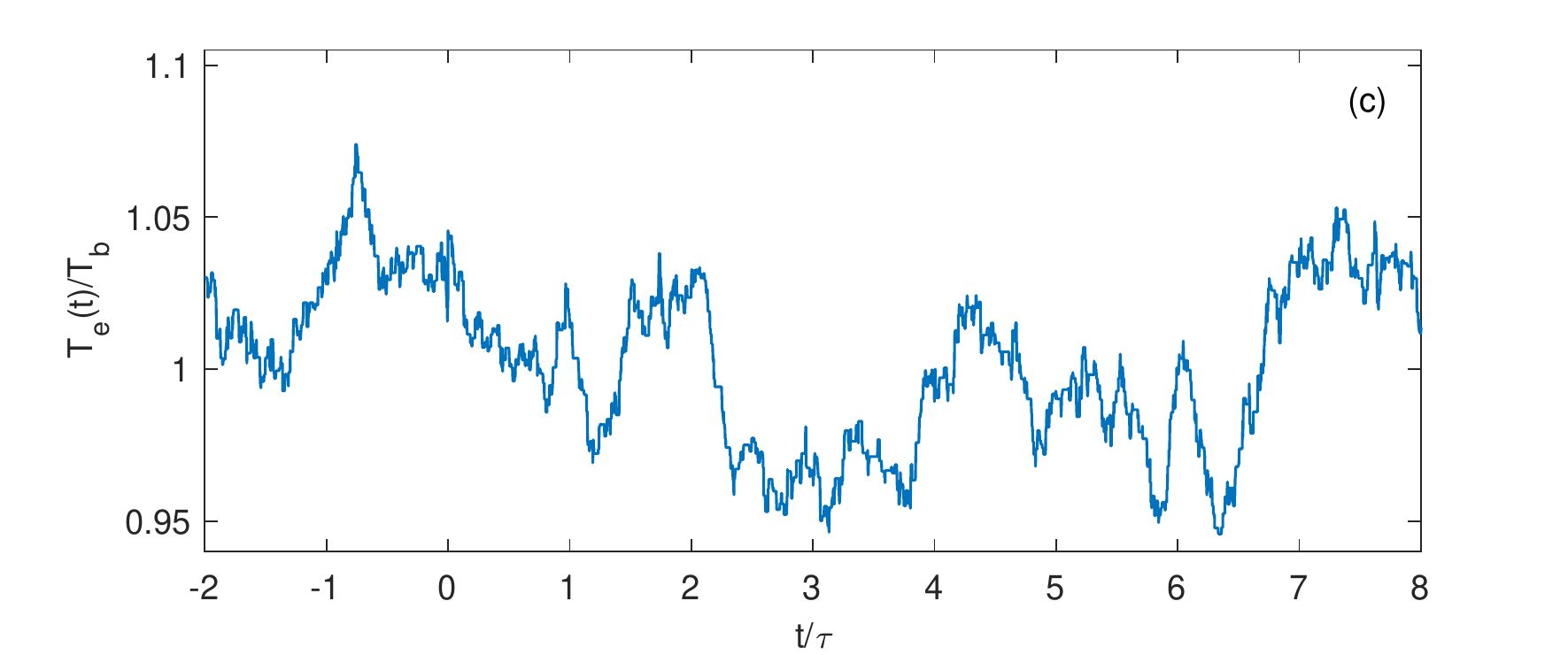}}
  \caption{Examples of Monte Carlo generated time traces of the temperature fluctuations for (a) $T_\text b = 5$~mK, (b) $T_\text b = 30$~mK and (c) $T_\text b = 100$~mK. Every time trace contains an injector event at $t=0$. In all cases, $C = 1000k_\text B$, $\varepsilon = 200$~$\mu$eV and $\tau$ denotes the relaxation time.}
\label{probdist}
\end{figure}

Now, evaluating the integrals explicitly, we obtain
\begin{eqnarray}
\nonumber
F_\text i(\xi_\text i,T_\text e) &=& \sqrt{\frac{2}{\pi}}g \bigg( K_1\left[\frac{\Delta}{k_\text BT_\text s}-i\xi_\text i \Delta\right]\cos\left[eV\xi_\text i\right]+ K_1\left[ \frac{\Delta}{k_\text BT_\text e}+i\xi_\text i \Delta \right] \cosh\left[\frac{eV}{k_\text BT_\text e}+ieV\xi_\text i\right]\\
&&\hspace{53mm}-K_1\left[\frac{\Delta}{k_\text BT_\text s}\right]- K_1\left[ \frac{\Delta}{k_\text BT_\text e}\right] \cosh\left[\frac{eV}{k_\text BT_\text e}\right] \bigg),
\end{eqnarray}
where $g=\frac{\sqrt{2\pi}G_\text T \Delta}{e^2}$ and $K_\text n[x]$ denotes the $n$th modified Bessel function of the second kind. Using that $k_\text BT\ll \Delta$, $T=T_\text s, T_\text e$, we simplify the Bessel functions as
\begin{equation}
K_1\left[\frac{\Delta}{k_\text BT} \pm i \xi_\text i\Delta\right] \approx \sqrt{\frac{\pi}{2}}h(T)e^{\mp i\xi_\text i\Delta},
\end{equation}
with $h(T) = \sqrt{\frac{k_\text BT}{\Delta}}e^{-\frac{\Delta}{k_\text BT}}$. This yields the following expression for the generating function for the injector-absorber junction:
\begin{equation}
F_\text i(\xi_\text i,T_\text e) = g\bigg( h(T_\text s)e^{i\xi_\text i\Delta} \cos\left[eV\xi_\text i\right]+h(T_\text e) e^{- i\xi_\text i\Delta} \cosh\left[\frac{eV}{k_\text BT_\text e}+ieV\xi_\text i\right]-h(T_\text s)-h(T_\text e)\cosh\left[\frac{eV}{k_\text BT_\text e}\right] \bigg).
\label{Main result injector}
\end{equation}

\subsubsection*{No applied bias (case I and II)}
For $V=0$, Eq.~\eqref{Main result injector} simplifies to
\begin{equation}
F_\text i(\xi_\text i,T_\text e) =g\left[ h(T_\text s)(e^{i\xi_\text i \Delta}-1)+ h(T_\text e) (e^{-i \xi_\text i \Delta}-1)\right].
\end{equation}
For $T_\text s \ll T_\text e$  ($T_\text s \gg T_\text e$), the second (first) term is negligible, yielding case (I) [(II)] in Eq.~(5) of the main text. In both cases, the statistics corresponds to Poissonian processes with an energy of $\Delta$ transferred in each elementary process.

\subsubsection*{Finite bias (case III)}
For $eV \gg kT_\text e$, we obtain from Eq.~\eqref{Main result injector}
\begin{equation}
F_\text i(\xi_\text i,T_\text e) = g\Bigg(\frac{h(T_\text s)}{2}\left[ e^{i(\Delta+eV)\xi_\text i}-1+e^{i(\Delta-eV)\xi_\text i}-1\right] +\frac{h(T_\text e)}{2} \left[e^{-i(\Delta-eV)\xi_\text i}-1\right]\Bigg).
\end{equation}
If $T_\text s (1-eV/\Delta) \ll T_\text e$, the first part is negligible and the cumulant generating function reduces to
\begin{equation}
F_\text i(\xi_\text i,T_\text e) =gh(T_\text e)e^{\frac{eV}{k_\text B T_\text e}}\left(e^{i(eV-\Delta)\xi_\text i}-1\right),
\end{equation}
which corresponds to case (III) in Eq.~(5) of the main text.

\subsection*{Generating function for the bath-absorber energy transfer}
At low temperatures, with a weak electron-phonon coupling, Fermi's golden rule yields the following counting field resolved rates
\begin{eqnarray}
\tilde \Gamma_{\pm}^\text b(\xi) \!\!\!&=& \!\!\!\frac{2\pi}{\hbar} \int dE_{\mathbf{k}} N_\text e(E_\mathbf{k}) f(E_\mathbf{k}) \int d\mathbf{q} N_\text b(\mathbf{q}) n_\pm(\varepsilon_\mathbf{q}) \mathcal{M}^2 \left[ 1-f(E_\mathbf{k\pm q})\right]\delta(E_\mathbf{k}-E_\mathbf{k\pm q}+\varepsilon_\mathbf{q}) e^{\pm i \varepsilon_\mathbf{q} \xi},
\label{Fermi's golden rule for electron-phonon coupling}
\end{eqnarray}
where $\tilde\Gamma_+(\xi)$ [$\tilde\Gamma_-(\xi)$] denotes the counting field resolved absorption (emission) rate of phonons, $E_\mathbf{k}$ ($\varepsilon_\mathbf{q}$) is the energy of an electron (phonon) with momentum $\mathbf{k}$ ($\mathbf{q}$), $N_\text e(\varepsilon)$ ($N_\text b(\varepsilon)$) is the density of states of electrons (phonons) on the island, $f(\varepsilon) = (\exp[\varepsilon/kT_\text e]+1)^{-1}$ is the Fermi function for the electrons, $n_+(\varepsilon) =(\exp[\varepsilon/(kT_\text b)]-1)^{-1}$ is the Bose distribution for the phonons, with $n_-(\varepsilon)=1+n_+(\varepsilon)$, and $\mathcal{M}$ is the coupling strength matrix element for electron-phonon scattering. The signs of the counting fields have been chosen such that positive energy corresponds to an inflow of energy to the electrons from the phonons. \\

\noindent At low temperatures, all relevant scattering processes occur around the Fermi level, i.e., $|\mathbf{k}|\approx |\mathbf{k}_\text F|$,  $|\mathbf{q}| \ll |\mathbf{k}_\text F|$ and $N(E_\mathbf{k}) \approx N_\text e$. We use a parabolic dispersion relation for the electrons in the metal, $E_\mathbf{k} = \frac{\hbar^2k^2}{2m} \equiv E_k$. Furthermore, the phonons are treated as longitudinal ones within the Debye model, i.e., $N_\text b(\mathbf{q}) = \mathcal{V}/(2\pi)^3 \equiv N_\text b$ and $\varepsilon_\mathbf{q} = \hbar c_l q\equiv \varepsilon_q$, where $c_l$ is the velocity of the phonons. For a scalar deformation potential, $\mathcal{M}^2 = \mathcal{M}_0^2q$ and Eq.~\eqref{Fermi's golden rule for electron-phonon coupling} can be written as
\begin{eqnarray}
\tilde \Gamma_{\pm}^\text b(\xi) \!\!\!&=& \!\!\!\frac{2\pi \mathcal{M}^2_0 N_\text e N_\text b}{\hbar} \int dE_{\mathbf{k}} f(E_k) \int d\mathbf{q} q \delta(E_k-E_\mathbf{k\pm q}\pm\varepsilon_q) \left[ 1-f(E_\mathbf{k\pm q})\right] n_\pm(\varepsilon_{q}) e^{\pm i \varepsilon_q \xi}.
\end{eqnarray}
Evaluating the integral over $\mathbf{q}$, we obtain
\begin{eqnarray}
\tilde \Gamma_{\pm}^\text b(\xi) \!\!\!&=& \!\!\!\frac{2\pi \mathcal{M}_0^2 N_\text e N_\text b}{\hbar^3c_\text l^2}\int dE_{\mathbf{k}} f(E_k) \int d\varepsilon \varepsilon^2 \frac{2\pi m}{\hbar^3 k_F c_l}\left[1-f(E_k\pm\varepsilon)\right]n_\pm(\varepsilon) e^{\pm i \varepsilon \xi} .
\end{eqnarray}
Now, we rewrite the integral as
\begin{eqnarray}
\tilde \Gamma_{\pm}^\text b(\xi) \!\!\!&=& \!\!\!\frac{(2\pi)^2m\mathcal{M}_0^2 N_\text e N_\text b}{\hbar^6c_\text l^3 k_\text F} \int d\varepsilon \varepsilon^2 n_\pm(\varepsilon) e^{\pm i \varepsilon \xi}\int dE  f(E)\left[1-f(E\pm\varepsilon)\right],
\end{eqnarray}
or
\begin{eqnarray}
\tilde \Gamma_{\pm}^\text b(\xi) \!\!\!&=& \!\!\!\frac{\mathcal{V}\mathcal{M}_0^2 N_\text e}{2\pi\hbar^5c_\text l^3 v_\text F} \int d\varepsilon \varepsilon^2 n_\pm(\varepsilon) e^{\pm i \varepsilon \xi}\int dE  f(E)\left[1-f(E\pm\varepsilon)\right],
\end{eqnarray}
where $v_\text F$ is the Fermi velocity of the electrons. The prefactor corresponds to $\Sigma \mathcal{V}/[24k_\text B^5\zeta(5)]$, while the integral over $E$ gives
\begin{equation}
\int^\infty_{-\infty} dE f(E) [1-f(E\pm\varepsilon)] = \varepsilon n_\mp(\varepsilon,T_\text e),
\end{equation}
where we have introduced a Bose distribution with explicit temperature dependence. We then obtain
\begin{eqnarray}
\tilde \Gamma_{\pm}^\text b(\xi) \!\!\!&=& \!\!\!\frac{\Sigma \mathcal{V}}{24 k_\text B^5 \zeta(5)} \int d\varepsilon \varepsilon^3 n_\pm(\varepsilon, T_\text b) n_\mp(\varepsilon,T_\text e) e^{\pm i \varepsilon \xi}.
\end{eqnarray}
The cumulant generating function is given by $F_\text b(\xi_\text b,T_\text b) = \Gamma_+^\text b(\xi_\text b)+\Gamma_-^\text b(\xi_\text b)-\Gamma_+^\text b(0)-\Gamma_-^\text b(0)$, or, equivalently,
\begin{equation}
F_\text b(\xi_\text b,T_\text b) = \int_0^\infty d\varepsilon \left[ \Gamma_+^\text b(\varepsilon) \left( e^{i\xi_\text b \varepsilon}-1\right)+\Gamma_-^\text b (\varepsilon) \left( e^{-i\xi_\text b \varepsilon}-1\right) \right],
\label{FCS phonon bath}
\end{equation}
with $\Gamma_{\pm}^\text b(\varepsilon) = \frac{\Sigma \mathcal{V}}{24k_\text B^5\zeta(5)}\varepsilon^3n_\pm(\varepsilon,T_\text b)n_\mp(\varepsilon,T_\text e)$. The cumulants are given by $S_\text b^{(n)} = \frac{\partial F_\text b(\xi_\text b,T_\text b)}{\partial \xi_\text b}\big|_{\xi_\text b  =0}$, yielding
\begin{equation}
S_\text b^{(n)} = \frac{\Sigma \mathcal{V}}{24k_\text B^5\zeta(5)} \int_0^\infty d\varepsilon \varepsilon^{3+n} \left[ n_+(\varepsilon,T_\text b)n_-(\varepsilon,T_\text e)\pm n_-(\varepsilon,T_\text b)n_+(\varepsilon,T_\text e)\right],
\end{equation}
with $+$ for $n$ even and $-$ for $n$ odd. For odd $n$, we obtain
\begin{eqnarray}
\nonumber
S_\text b^{(n)}  &=&  \frac{\Sigma \mathcal{V}}{48k_\text B^5\zeta(5)} \int_0^\infty d\varepsilon \varepsilon^{3+n} \left[\coth\left(\frac{\varepsilon}{2k_\text B T_\text b}\right)-\coth\left(\frac{\varepsilon}{2k_\text B T_\text e}\right)\right]\\
&=&  \Sigma\mathcal{V}k_\text B^{n-1}\frac{\zeta(n+3)(n+3)!}{24\zeta(5)}\left(T_\text b^{n+4}-T_\text{e}^{n+4} \right), \qquad n = 1,3, 5, ...
\end{eqnarray}
while for even $n$, we obtain
\begin{eqnarray}
\nonumber
S_\text b^{(n)}  &=&  \frac{\Sigma \mathcal{V}}{48k_\text B^5\zeta(5)} \int_0^\infty d\varepsilon \varepsilon^{3+n} \left[\coth\left(\frac{\varepsilon}{2k_\text B T_\text b}\right)\coth\left(\frac{\varepsilon}{2k_\text B T_\text e}\right)-1\right]\\
& \approx& \Sigma\mathcal{V}k_\text B^{n-1}\frac{\zeta(n+4)(n+3)!}{24\zeta(5)}\left(T_\text b^{n+4}+T_\text{e}^{n+4}\right), \qquad n = 2,4,6,...
\end{eqnarray}
In the last step, we have made use of the following approximation:
\begin{equation}
I_1 \equiv \int_0^{\infty}d\varepsilon ~\varepsilon^{3+n} \left[\coth(\varepsilon r)\coth(\varepsilon)-1\right] \approx \int_0^{\infty}d\varepsilon \frac{\varepsilon^{3+n}}{2}\left[\coth^2(\varepsilon)-1\right]\left(1+\frac{1}{r^6}\right) \equiv I_2.
\end{equation}
To estimate the accuracy of this approximation, we first perform a change of variables $\varepsilon \rightarrow \varepsilon r$ in the second term in $I_2$ to obtain
\begin{equation}
I_2=\int_0^{\infty}d\varepsilon ~\varepsilon^{3+n} \left[\frac{\coth^2(\varepsilon)+ \coth^2(\varepsilon r)}{2}-1\right]
\end{equation}
with which we get
\begin{equation}
I_2-I_1=\int_0^{\infty}d\varepsilon ~ \frac{\varepsilon^{3+n}}{2}\left[\coth(\varepsilon)-\coth (\varepsilon r)\right]^2.
\end{equation}
By noting that $\coth(\varepsilon) \geq \coth(\varepsilon r) \geq 1$ for any $\varepsilon \geq 0$ and $r \geq 1$, we have that 
\begin{equation}
\frac{I_2-I_1}{I_2}\leq \frac{\int_0^{\infty}d\varepsilon ~ \varepsilon^{3+n}\left[\coth(\varepsilon)-1\right]^2}{\int_0^{\infty}d\varepsilon ~ \varepsilon^{3+n}\left[\coth^2(\varepsilon)-1\right]}\leq \frac{\int_0^{\infty}d\varepsilon ~ \varepsilon^{5}\left[\coth(\varepsilon)-1\right]^2}{\int_0^{\infty}d\varepsilon ~ \varepsilon^{5}\left[\coth^2(\varepsilon)-1\right]}=1-\frac{\pi^6}{945 \zeta(5)}\approx 0.0189
\end{equation}
with the first inequality becoming an equality only for $r\rightarrow \infty$.

\subsection*{Stochastic path integral formulation}

The starting point for the derivation of the full statistics of the time-integrated temperature fluctuations $\theta=\int_0^{t_0}dt[T_\text e(t)-\overline T_\text e]$ is the generating functions for energy transfers between the injector and the absorber, $\Delta t F_\text i [\xi_\text i (t),T_\text e(t)]$, and the bath and the absorber, $\Delta t F_\text b[\xi_\text b(t),T_\text e(t)]$, during a time interval $[t,t+\Delta t]$. The length of the time interval $\Delta t$ is so short that the absorber temperature is only marginally changed, $T_\text e(t+\Delta t)\approx T_\text e(t)+\Delta T_\text e(t)$, where $\Delta T_\text e(t) \ll T_\text e(t)$. This requires $\Delta t$ to be much shorter than the time scale over which $T_\text e(t)$ changes appreciably, typically set by $\tau$.

In an interval $\Delta t$, for transferred energies $\Delta E_\text i$ and $\Delta E_\text b$, the corresponding energy currents are $I_{\text{Ei}}=\Delta E_\text i/\Delta t$ and $I_{\text{Eb}}=\Delta E_\text b/\Delta t$, for the injector-absorber and bath-absorber transfers respectively. For the entire measurement time $t_0$, taking the continuum-in-time limit, we can write the joint, unconditioned probability distribution of energy curents as a product of the individual probabilities as
\begin{equation}
P[I_\text{Ei},I_\text{Eb}]=P[I_\text{Ei}]P[I_\text{Eb}]
\label{uncon}
\end{equation}
where the probabilities $P[I_\text{Ei}]$, $P[I_\text{Eb}]$ conveniently can be written as stochastic path integrals as
\begin{equation}
P[I_\text{Ei}]=\int {\mathcal D} [\xi_\text i]\exp \left[\int_0^{t_0} dt\left(-i I_\text{Ei}(t)\xi_\text i (t)+F_\text i [\xi_\text i (t),T_\text e(t)]\right)\right],
\label{unconi}
\end{equation}
and
\begin{equation}
P[I_\text{Eb}]=\int {\mathcal D} [\xi_\text b]\exp \left[\int_0^{t_0} dt\left(-i I_\text{Eb}(t)\xi_\text b (t)+F_\text b [\xi_\text b (t),T_\text e(t)]\right)\right].
\label{unconb}
\end{equation}
To account for the effect of the transferred energy, with resulting fluctuations of $T_\text e(t)$, and following back-action on the statistics on the transfer events themselves, we have the absorber energy $E(t)$ conservation equation
\begin{equation}
\frac{dE(t)}{dt}=I_\text{Ei}(t)+I_\text{Eb}(t).
\end{equation}
Importantly, the total energy of the absorber is direxctly related to the temperature via the relation $E(t)=C[T_\text e(t)] T_\text e(t)/(2C)$ with $C[T_\text e(t)]\propto T_\text e(t)$. The conditioned probability for the realizations of the energy currents is then given by the unconstrained one multiplied by a functional $\delta$-function as
\begin{equation}
P[I_\text{Ei},I_\text{Eb}]\delta\left[\frac{dE(t)}{dt}-I_\text{Ei}(t)+I_\text{Ei}(t)\right]
\end{equation}
Integrating the constrained probability over the energy currents we get, writing the $\delta$-function as a functional Fourier transform and inserting the expression in Eq. (\ref{uncon}), 
\begin{equation}
\int {\mathcal D}[I_\text{Ei}] {\mathcal D}[\xi_\text i] {\mathcal D}[I_\text{Eb}] {\mathcal D}[\xi_\text b] {\mathcal D}[\xi] \exp \left[\int_0^{t_0} dt H(t)\right]
\end{equation}
where $H(t)=H[t,I_\text{Ei}(t) \xi_\text i (t),I_\text{Eb}(t) \xi_\text b (t),\xi(t)]$ is
\begin{eqnarray}
H(t)&=&i\xi(t) \left(\frac{dE(t)}{dt}-I_\text {Ei}(t)- I_\text {Eb}(t)\right) \nonumber \\
&-&iI_\text {Ei}(t)\xi_\text i (t)+F_\text i[\xi_\text i(t),T_\text e(t)]-iI_\text {Eb}(t)\xi_\text b (t)+F_\text b[\xi_\text b(t),T_\text e(t)]
\end{eqnarray}
We can now perform the integrals over $I_\text{Ei}(t)$ and  $I_\text{Eb}(t)$, giving functional delta functions $\delta[\xi_\text i(t)-\xi(t)]$ and  $\delta[\xi_\text b(t)-\xi(t)]$ and hence the total, constrained probability
\begin{equation}
\int {\mathcal D}[\xi] \exp \left[\int_0^{t_0} G[t,\xi(t),T_\text e(t)]\right]
\end{equation}
where
\begin{eqnarray}
G[t,\xi(t),T_\text e(t)]&=&i\xi(t) \frac{dE(t)}{dt}+F_\text i[\xi(t),T_\text e(t)]+ F_\text b[\xi(t),T_\text e(t)]
\end{eqnarray}
This expression thus gives the probability distribution of realizations of the total energy change, $dE(t)/dt$. To access the statistics of the realizations of the temperature we conveneintly multiply the obtained probability distribution by a delta function $\delta[T(t)-T_\text e(t)]$, recalling the relation between $E(t)$ and $T_\text e(t)$, and integrate over $E(t)$ giving
\begin{eqnarray}
P[T]=\int {\mathcal D}[\chi] \exp \left[\int_0^{t_0} dt\left(-i\chi(t)T(t)+\lambda[t,\chi(t)]\right)\right]
\end{eqnarray}
where  
\begin{eqnarray}
&&\exp \left[\int_0^{t_0} dt \lambda[t,\chi(t)]\right] = \int {\mathcal D}[\xi]{\mathcal D}[E]\exp \left[\int_0^{t_0} \left(i\chi(t)T_\text e(t)+G[t,\xi(t),T_\text e(t)]\right)\right]
\end{eqnarray}
is a stochastic path integral over $\xi(t),E(t)$.

\subsubsection*{Long time limit}
In the limit of a long measurement time $t_0$ we can neglect the time dependence of the variables and write the probability distribution of the time-integrated temperature $\theta=\int_0^{t_0}[T_\text e(t)-\overline T_\text e]dt$ as (up to phase factor shifting the distribution) 
\begin{equation}
P(\theta)=\frac{1}{2\pi} \int d\chi \exp \left[-i\chi \theta+\lambda(\chi)\right]
\end{equation}
where
\begin{eqnarray}
e^{\lambda(\chi)}=\int d\xi dE \exp \left[t_0 S(\chi,\xi,T_\text e)\right]
\end{eqnarray}
and
\begin{eqnarray}
S(\chi,\xi,T_\text e)=i\chi (T_\text e-\overline T_\text e)+ F_\text i[\xi,T_\text e]+ F_\text b[\xi,T_\text e]
\end{eqnarray}
Solving this equation in the saddle point approximation we get the generating function, to exponential accuracy, as 
\begin{equation}
\lambda(\chi)=t_0S(\chi,\xi^*,T_\text e ^*)
\end{equation}
where $\xi^*=\xi^*(\chi)$ and $T_\text e^*=T_\text e^*(\chi)$ are the solutions of the saddle point equations
\begin{eqnarray}
\frac{\partial S}{\partial \xi}&=&\frac{\partial F_\text i}{\partial \xi}+\frac{\partial F_\text b}{\partial \xi}=0 \nonumber \\
\frac{\partial S}{\partial E}&\propto& \frac{\partial S}{\partial T_\text e}= i\chi+\frac{\partial F_\text i}{\partial T_\text e}+\frac{\partial F_\text b}{\partial T_\text e}    =0 
\label{Saddle point eqs}
\end{eqnarray}
From Eq.~\eqref{Saddle point eqs} and $\lambda(\chi)$ we obtain the low-frequency cumulants of the temperature fluctuations as $S_{\text T_\text e}^{(n)}=(1/t_0)(-i)^n\partial^n_{\chi} \lambda(\chi)|_{\chi=0}$. In terms of $\langle \! \langle \mathcal{E}^n (T_\text e)\rangle \! \rangle=(-i)^n\partial_{\xi}^n F(\xi,T_\text e)|_{\xi=0}$, the cumulants of the absorber energy currents,  the average temperature $\overline T_\text e$ is found from  $\langle \mathcal{E} (\overline T_\text e)\rangle=0$, yielding the equation
\begin{equation}
h(T_\text s)+h(\overline T_\text e) \left[-\cosh\left( \frac{eV}{k_\text B\overline T_\text e}\right)+\frac{eV}{\Delta}\sinh\left(\frac{eV}{k_\text B\overline T_\text e}\right)\right] = \frac{1}{5r}\left(\frac{\overline T_\text e^5}{T_\text b^5}-1\right),
\label{Average temperature}
\end{equation}
where $h(T) = \sqrt{\frac{k_\text BT}{\Delta}}e^{-\frac{\Delta}{k_\text BT}}$ as before and $r=\frac{\sqrt{2\pi}G_\text T\Delta^2}{T_\text b e^2 \kappa}$. The second and third temperature cumulants, experimentally most relevant,  are given by
\begin{eqnarray}
\nonumber
S_{\text T_\text e}^{(2)}&=&\frac{1}{\kappa^2}\langle \! \langle \mathcal{E}^2(T_\text e) \rangle \! \rangle , \\
S_{\text T_\text e}^{(3)}&=&\frac{1}{\kappa^3}\left[\langle \! \langle \mathcal{E}^3(T_\text e) \rangle \! \rangle+3\langle \! \langle \mathcal{E}^2(T_\text e) \rangle \! \rangle \frac{d}{dT_\text e}\frac{ \langle \! \langle \mathcal{E}^2(T_\text e) \rangle \! \rangle}{\kappa (T_\text e)}\right],
\label{Energy temperature eqs}
\end{eqnarray}
where $\kappa (T_\text e)=i\partial_{T_\text e}\partial_\xi F(\xi,T_\text e)|_{\xi=0}$, the heat conductance and all quantities in Eq. \eqref{Energy temperature eqs} are evaluated at $\overline T_\text e$. This is Eq.~(9) of the main text.

Of particular interest is the regime $\tau \ll 1/\Gamma_\text i$, with well separated energy injection events. Then $\overline T_\text e \approx T_\text b +\Delta T$, with $\Delta T=\Gamma_\text i \langle \varepsilon\rangle/\kappa$ and $\kappa\equiv \kappa (T_\text b)$, deviates negligiably from $T_\text b$.
The temperature noise $S_{\text T_\text e}^{(2)}$ in Eq. \eqref{Energy temperature eqs} becomes, to leading order in $\Delta T/T_\text b \ll 1$,
\begin{eqnarray}
\frac{S_{\text T_\text e}^{(2)}}{S_{0}^{(2)}}= \frac{1}{2z^2}\left[1+\left(\frac{\overline T_\text e}{T_\text b}\right)^6\right]+\frac{r\beta}{2z^2}\left[h(T_\text s)\left[1+\left(\frac{eV}{\Delta}\right)^2\right]+h(\overline T_\text e)H(\overline T_\text e,V) \right],
\label{avnoise}
\end{eqnarray}
where $S_0^{(2)}=\frac{2k_\text BT_\text b^2}{\kappa}$, $\beta = \frac{\Delta}{k_\text B T_\text b}$,  $H(T,V) = \left[1+\left(\frac{eV}{\Delta}\right)^2\right]\cosh \left(\frac{eV}{k_\text BT}\right)-2\frac{eV}{\Delta}\sinh\left(\frac{eV}{k_\text B T}\right)$ and
\begin{equation}
 z \equiv \frac{\kappa(T_\text e)}{\kappa} = \left(\frac{\overline T_\text e}{T_\text b} \right)^4+r\beta\left(\frac{T_\text b}{\overline T_\text e}\right)^2h(\overline T_\text e)H(\overline T_\text e,V).
\end{equation}
For only thermal bias, we obtain from Eq.~\eqref{Average temperature}
\begin{equation}
\Delta T  = r T_\text b\left(h(T_\text s)+h(T_\text b)\left[-\cosh\left( \frac{eV}{k_\text B T_\text b}\right)+\frac{eV}{\Delta}\sinh\left(\frac{eV}{k_\text B T_\text b}\right)\right] \right).
\end{equation}
Furthermore, $H(\overline T_\text e,V)  = 1$. If $\beta \gg \ln(r)\gg 1$, we have $z = q^4$, where $q = \frac{\overline T_\text e}{T_\text b}$. The normalized second cumulant in Eq.~\eqref{avnoise} then reduces to
\begin{equation}
\frac{S_{\text T_\text e}^{(2)}}{S_{0}^{(2)}}= \frac{1+q^6+(\beta/5)[q^5-1]}{2\tau^8}
\end{equation}
which is Eq.~(10) of the main text.

For voltage bias only, $T_\text s = T_\text b$, and $rh(T_\text b) \ll 1$, Eq.~\eqref{Average temperature} reduces to
\begin{equation}
e^{-(\Delta -eV)/[k_\text B \overline T_\text e]} = \frac{2}{5r}\frac{\Delta ^{3/2}}{\sqrt{\overline T_\text e}(\Delta -eV)}\left(1-\frac{\overline T_\text e^5}{T_\text b^5}\right).
\end{equation}
Furthermore, we have $z=q^4+\frac{\tilde{\beta}_\text b(1-q^5)}{5q^2}$, where $\tilde{\beta}=\beta(1-\frac{eV}{\Delta})$. The normalized second cumulant in Eq.~\eqref{avnoise} then reduces to
\begin{equation}
\frac{S_{\text T_\text e}^{(2)}}{S_{0}^{(2)}}= \frac{q^4}{2}\frac{1+q^6+(\beta/5)[1-q^5]}{\left(q^6+(\tilde{\beta}_\text b/5)(1-q^5)\right)^2},
\end{equation}
which is Eq.~(11) of the main text.